\documentclass{aastex631}
\usepackage{amssymb, amsmath,framed}
\usepackage{gensymb}

\usepackage{bm}
\expandafter\ifx\csname package@font\endcsname\relax\else
 \expandafter\expandafter
 \expandafter\usepackage
 \expandafter\expandafter
 \expandafter{\csname package@font\endcsname}

\fi
\def\bq{\begin{equation}}
\def\eq{\end{equation}}
\def\bqy{\begin{eqnarray}}
\def\eqy{\end{eqnarray}}

\def\p{\partial}

\def\p{\partial}

\def\R{\mathbb{R}}

\submitjournal{the Astronomical Journal}

\begin{document}
\title{\large{Interstellar Objects in the Solar System:
\\1. Isotropic Kinematics from the Gaia Early Data Release 3}}

\correspondingauthor{T. Marshall Eubanks}
\email{tme@space-initiatives.com}

\author{T. Marshall Eubanks}
\affiliation{Space Initiatives Inc, Newport, VA 24128, USA}
\affiliation{Institute for Interstellar Studies - US (i4is-US)}

\author{Andreas M. Hein}
\affiliation{Initiative for Interstellar Studies (i4is) 27/29 South Lambeth Road London, SW8 1SZ United Kingdom}

\author{Manasvi Lingam}
\affiliation{Department of Aerospace, Physics and Space Sciences, Florida Institute of Technology, Melbourne FL 32901, USA}
\affiliation{Institute for Theory and Computation, Harvard University, Cambridge MA 02138, USA}

\author{Adam Hibberd}
\affiliation{Initiative for Interstellar Studies (i4is) 27/29 South Lambeth Road London, SW8 1SZ United Kingdom}

\author{Dan Fries}
\affiliation{Department of Aerospace Engineering and Engineering Mechanics, University of Texas at Austin, Austin TX 78712, USA}
\affiliation{Initiative for Interstellar Studies (i4is) 27/29 South Lambeth Road London, SW8 1SZ United Kingdom}

\author{Nikolaos Perakis}
\affiliation{Initiative for Interstellar Studies (i4is) 27/29 South Lambeth Road London, SW8 1SZ United Kingdom}
\affiliation{ Department of Space Propulsion, Technical University of Munich, Germany}

\author{Robert Kennedy}
\affiliation{Institute for Interstellar Studies - US (i4is-US)}

\author{W. P. Blase}
\affiliation{Space Initiatives Inc, Newport, VA 24128, USA}

\author{Jean Schneider}
\affiliation{Observatoire de Paris - LUTH, 92190 Meudon, France}

\begin{abstract}

1I/'Oumuamua (or 1I) and 2I/Borisov (or 2I), the first  InterStellar Objects (ISOs) discovered passing through the solar system, have opened up entirely new areas of exobody research. Finding additional ISOs and planning missions to intercept or rendezvous with these bodies
will greatly benefit from knowledge of their likely orbits and arrival rates.
Here, we use
the local velocity distribution of stars 
from the Gaia Early Data Release 3 Catalogue of Nearby Stars and a standard gravitational focusing model to predict the velocity dependent flux of ISOs entering the solar system. With an 1I-type ISO number density of $\sim$0.1 AU$^{-3}$, we predict that a total of $\sim$6.9 such objects per year should pass within 1 AU of the Sun. There will be a fairly large high-velocity tail to this flux, with half of the  incoming ISOs  predicted to have  a velocity at infinity, v$_{\infty}$, $>$ 40 km s$^{-1}$. Our model predicts that $\sim$92\% of incoming ISOs will be residents of the galactic thin disk, $\sim$6\% ($\sim$4 per decade) will be from the  thick disk, $\sim$1 per decade will be from the  halo and at most  $\sim$3 per century will be unbound objects, ejected from our galaxy or entering the Milky Way from another galaxy.  The rate of ISOs with very low v$_{\infty}$ $\lesssim$ 1.5 km s$^{-1}$ is so low in our model that any incoming very low velocity ISOs are likely to be  previously lost solar system objects. Finally, 
we estimate a cometary ISO number density of $\sim$7 $\times$ 
10$^{-5}$ AU$^{-3}$ for 2I type ISOs, leading to discovery rates for these objects possibly approaching once per decade with future telescopic surveys.

\end{abstract}

\section{Introduction} \label{SecIntro}

Interstellar objects (ISOs) passing through the solar system can be directly observed by Earth-based telescopes and potentially explored at close range by spacecraft. 
Because galactic dynamics mixes material from different parts of the Galaxy, the
direct \textit{in situ} exploration of ISO will enable the direct sampling of different regions of the galaxy, and of their history.
Missions to passing exobodies will truly be interstellar missions, providing scientific returns that would take millennia or longer to obtain with even fast interstellar travel. ISOs will originate throughout the birth, life and death of stellar and planetary systems \citep{Eubanks-2019-c}, and  
can be expected to share the galactic velocities of their original systems, possibly dispersed by their ejection from their original host systems \citep{Siraj-Loeb-2020-c}.

Any interstellar object passing through the solar system  will be on a parabolic or hyperbolic orbit relative to the Sun, with a velocity ``at infinity'' (i.e., far from the Sun), v$_{\infty}$,   $\geq$ 0, and an eccentricity, e, $\geq$ 1.  
The recent discovery of the first two such ISOs
opens the potential for the direct observation of these exobodies, both telescopically 
(see, e.g., \citep{Trilling-et-al-2018-b,Guzik-et-al-2019-a}, and with flyby, rendezvous and sample return spacecraft missions \citep{Seligman-Laughlin-2018-a,Hein-et-al-2017-a,Hein-et-al-2020-b}. These missions 
can provide direct, \textit{in situ} observatons on the shape, density, composition, isotopic abundances, and galactic history of ISOs \citep{Eubanks-et-al-2020-b,Moore-et-al-2020-a}. 
The amazing diversity of the planetary systems being found in exoplanetary research \citep{Winn-Fabrycky-2015-a,Perryman-2018-a,Lingam-Loeb-2021-a}
strongly suggests that there will a corresponding diversity in the ISOs passing through the solar system, especially considering ISOs will not just result from ejection from protoplanetary 
disks \citep{Portegies-Zwart-et-al-2017-a,Moro-Martin-2018-a,Moro-Martin-2018-b,Hands-Dehnen-2020-a}, 
but also from processes during \citep{Portegies-Zwart-2020-a,Zhang-Lin-2020-a} and 
after \citep{Eubanks-2019-c} the main sequence lifetime of planetary systems, and even from
the disruption of bodies in 
 white dwarf \citep{Rafikov-2018-b} or pulsar systems \citep{Brook-et-al-2014-a}.
A long term program to find and explore ISOs can potentially sample a wide range of these types with 
current technology, decades or even centuries before comparable missions will reach even the nearest stars.

The feasibility of reaching interstellar objects passing through the solar system has been assessed in
\cite{Seligman-Laughlin-2018-a}, with specific missions to the interstellar objects 1I/'Oumuamua and 2I/Borisov being presented by 
\cite{Hein-et-al-2017-a}  and \cite{Hibberd-et-al-2019-b}. Such missions are feasible 
with more massive spacecraft (100 kg or larger) using existing rockets and technologies
either if they can be initiated around the time of perihelion passage, e.g. by a Comet Interceptor type mission \citep{Schwamb-et-al-2020-a}, or by using a combination of planetary flybys and solar Oberth maneuvers (rocket accelerations at high speed close to the Sun) to overtake 
the ISO as it retreats from the Sun \citep{Hein-et-al-2019-a,Hein-et-al-2017-a}.
 
In an earlier paper \citep{Hein-et-al-2020-b} the authors introduced a 9-type ISO taxonomy, each type being based on populations observed in the galaxy or expected in the solar system, in order to assist the planning of missions and observations. Types 1 - 3 are based on the well known structural components of spiral galaxies such as the Milky Way, as defined kinematically in \cite{Gaia-et-al-2020-b}, type 4 was added to include objects not gravitationally bound to our galaxy and type 5 to distinguish very slow objects  as these are likely to be escapees from the Oort cloud re-encountering the solar system. Types 6-9 describe ISOs captured into different solar system orbits, and are not covered in this paper. Section \ref{Sec:Stellar-Velocity-Distributions} describes these types further and Table \ref{table:kinematic-stats} provides statistics of their stellar populations using data from the Gaia EDR3 GCNS \citep{Gaia-et-al-2020-c,Gaia-et-al-2020-b}. Section \ref{Sec:Arrival-Rate-Estimates} describes how these data were used to estimate the ISO arrival flux as a function of velocity 

Almost 95\% of the GCNS velocity data set stars are from the galactic thin disk, including both 1I and 2I. While this population  will undoubtedly be the most common type of ISO arriving in the solar system,  objects from other populations are also present in the solar neighborhood. Type 2 ISOs were thus defined to include the thick disk, with typical velocities relative to the Sun of $\gtrsim$100 km s$^{-1}$, and type 3 ISOs, galactic halo objects, are defined to have velocities relative to the Sun $\gtrsim$200 km s$^{-1}$ \citep{Nissen-Schuster-2010-a}. In general these other kinematic types will consist of older objects. The  type 2 thick disk stars are predominately over 8 billion years old while the type 3 Halo objects appear to be a complicated mixture of stars acquired in previous galactic mergers, stars ejected from the galactic disk, and stars, potentially very old, that formed in the halo  \citep{Haywood-et-al-2013-a,Johnston-2016-a}. All of these populations can be expected to contribute ISOs to the population arriving at the solar system, although the number density of ISOs from a given stellar population may depend on stellar age and metallicity. The Gaia data also 
show a population of gravitationally unbound stars passing through the galactic   \citep{Marchetti-2020-a}, enabling the prediction of the arrival rate of 
type 4 ISOs, which includes any bodies not gravitationally bound to the Milky Way galaxy. Type 4 ISOs are probably dominated by  
 objects being ejected from our galaxy  \citep{Marchetti-2020-a}, but could also include objects arriving here from other galaxies. Although the high velocity type 2, 3 and 4 objects will be difficult targets for spacecraft missions,  they would also be very rewarding sources of scientific data, e.g., on the formation and history of the galaxy. Finally, type 5 objects are 
 in galactic orbits, not 
 bound to the Sun but with  v$_{\infty}$ $\leq$ 1.5 km s$^{-1}$ relative to the solar system. This type was added as external ISOs with
 very low relative velocities are likely to be greatly outnumbered by  objects from the ``Oort spike'' (the sharp peak in the velocity distribution function for incoming long period comets with semi-major axes $>$ 10$^{4}$ AU) \citep{Krolikowska-et-al-2013-a}. In practice, Type 5 ISos will include both bound objects that appear to be unbound due to perturbations, and  weakly-bound Oort cloud objects that have escaped the Sun's gravity and are now re-encountering the solar system.

A substantial fraction of the stars in the solar neighborhood are concentrated 
in the collections
known as dynamical streams, associations or moving groups \citep{Famaey-et-al-2005-a,Kushniruk-et-al-2017-a,Gaia-et-al-2018-b}. Both 1I and 2I have been linked to dynamical streams; 1I appears to be part of the dynamically young Pleiades stream 
\citep{Feng-Jones-2018-a,Eubanks-2019-a,Eubanks-2019-b} while 2I may have been a member of the older, smaller (and higher metallicity) Wolf 630 stream \citep{Eubanks-2019-c}. This paper will concentrate on isotropic models for the ISO velocities and incoming flux; our subsequent paper will examine the relations between ISOs and the galactic dynamics revealed in the Gaia EDR-3.

\section{Gravitational Focusing Of Incoming Interstellar Objects}
\label{Sec:Grav-Focusing}

\begin{figure}[ht!]
\plotone{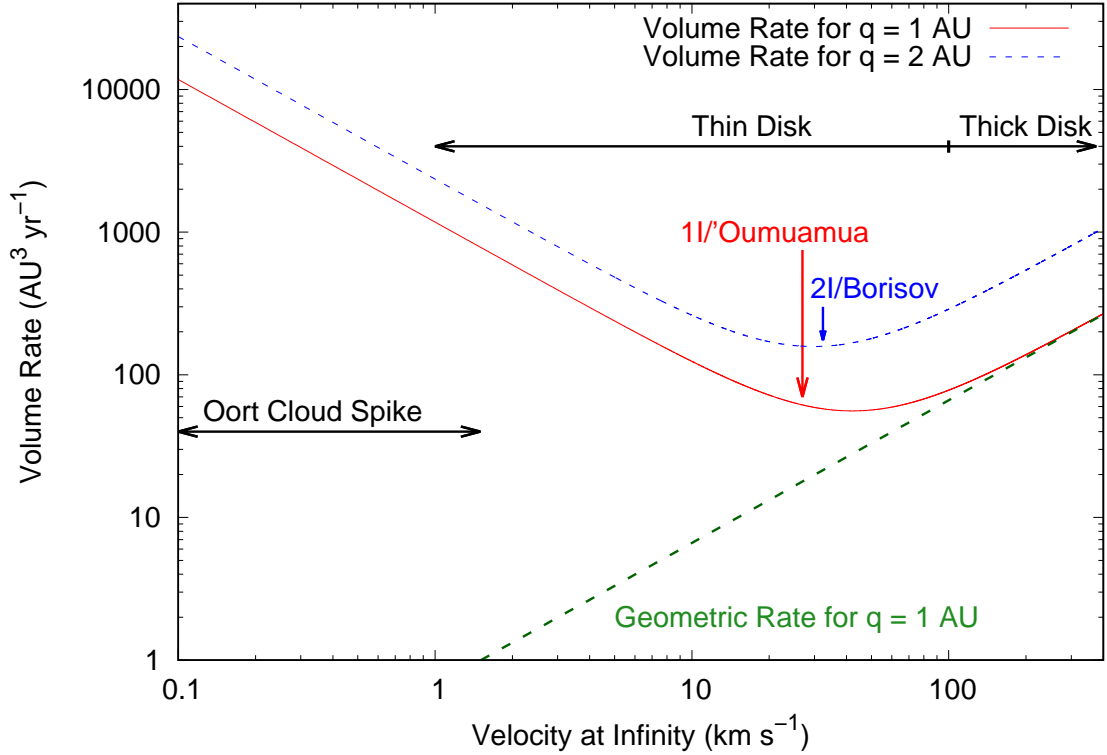}
\caption{The volume sampling rate, $\gamma(\mathrm{v}_{\infty},\mathrm{q})$, as a function of the velocity at infinity, v$_{\infty}$, after accounting for gravitational focusing via Equations \ref{eq:cross-section} and \ref{eq:gamma-normalized}. 
The dashed line shows the geometric rate, the volume sampling rate for q = 1 AU in the absence of any gravitational forcing. While the geometric rate is adequate for velocities $\gg$ the local solar escape, in the inner solar system gravitational focusing dominates the slower moving part of the ISO velocity distribution. While 1I and 2I have v$_{\infty}$ near the minimum of the volume sampling rate here, Figure \ref{fig:PDF-velocity} shows that this is misleading; because of the likely ISO velocity distributions the v$_{\infty}$ for the two observed ISOs are  near the maximum of the estimated ISO arrival rate.
\label{fig:volume_rate}
}
\end{figure}

The Sun's gravity deflects incoming unbound particles towards the Sun, increasing their density and velocity, a phenomenon known as gravitational focusing. 
For a given velocity at infinity, 
v$_{\infty}$, and perihelion, q, the gravitational focusing cross section, $\sigma$, for an object of negligible mass is 
\citep{Raymond-et-al-2017-a}
\begin{equation}
\sigma(\mathrm{v}_{\infty},\mathrm{q})\ =\ \pi\ \mathrm{q}^{2}\ \left[1 + \left(\frac{\mathrm{v}_{esc}(\mathrm{q})}{\mathrm{v}_{\infty}}\right)^{2}\right]
\label{eq:cross-section}
\end{equation}
where v$_{esc}$ is the solar escape velocity at the perihelion distance, q, given by
\begin{equation}
\mathrm{v}_{esc}(\mathrm{q})\ =\ \sqrt{\frac{2\ \mathrm{G}\mathrm{M}_{\bigodot}}{q}}
\label{eq:v_escape}
\end{equation}
where G is the gravitational constant and M$_{\bigodot}$ the solar mass.
The normalized volume sampling rate for an isotropic flux for a given q is simply
\begin{equation}
\gamma(\mathrm{v}_{\infty},\mathrm{q})\ =\ \mathrm{v}_{\infty}\ \sigma(\mathrm{v}_{\infty},\mathrm{q})
\label{eq:gamma-normalized}
\end{equation}
Figure \ref{fig:volume_rate} shows this rate as a function of v$_{\infty}$.
In order to estimate the isotropic ISO flyby rate, it is also necessary to have an estimate for the ISO number density and an estimate of the ISO velocity probability distribution as a function of incoming  v$_{\infty}$.

\section{The Number Density of Small Interstellar Objects}
\label{Sec:1I-Number-Density}

The first ISO known to visit
our solar system was discovered on October 19, 2017. This object, named 1I/'Oumuamua, was detected, tracked, and observed  as it
was moving through the solar system at a heliocentric velocity of
~50 km/s. \textit{Pan-STARRS1} detected 'Oumuamua after $\sim$3.5 years of observing in its current survey mode, which  \cite{Do-et-al-2018-a} used to calculate an upper limit of $\sim$0.2 AU$^{-3}$ to the space density, n$_{\mathrm{ISO}}$, of similar sized ISOs. Given that the observational duration has roughly doubled since then, and that surveys continue to improve, we halve the \cite{Do-et-al-2018-a} estimate, and adopt
\begin{equation}
\mathrm{n}_{\mathrm{ISO}}\ \lesssim\ 0.1\ \mathrm{AU}^{-3}\ .
\label{eq:number-density}
\end{equation}
for an upper bound of the number density of 1I sized ISOs. This estimate of n$_{\mathrm{ISO}}$, together with the known gravitational focusing of the solar system, and an model for the ISO velocity distribution, is needed to estimate of the differential arrival 
rate, $\Gamma$(v$_{\infty}$,q), where q is the perihelion of the incoming hyperbolic orbit. 
This paper will derive an isotropic $\Gamma$ estimate assuming an isotropic velocity distribution; a subsequent paper will derive directional velocity distributions and relate these to the the dynamics and resonances of the galaxy in the solar neighborhood.

\section{The Velocity-Number Distribution of Nearby Stars}
\label{Sec:Stellar-Velocity-Distributions}

While the ISO number density as a function of  v$_{\infty}$ is 
presently poorly unconstrained, it is reasonable to assume that the normalized velocity distribution of ISOs, p$_{ISO}$(v) is close to the \textit{stellar} velocity 
distribution, p$_{\mathrm{*}}$(v) in the solar neighborhood.  

We use the Gaia EDR-3 GCNS 3-D velocity sample to determine p$_{\mathrm{GCNS}}$(v) and use that as a proxy for p$_{ISO}$(v).
The 331,312 stars in the GCNS  are thought to include at least 92\% of stars of stellar type M9 or earlier within 100\,pc of the Sun, providing a nearly complete catalog of stars within the solar neighborhood \citep{Gaia-et-al-2020-b}. However, due to a lack of radial velocities we could only use a total of 77,132 stars from this catalog; these stars have the radial velocity  data needed to provide all three components of 3-D velocity, and also pass two catalog quality checks, requiring the ``probability of having reliable astrometry'' to be $\ge$ 0.75 and the ``maximum renormalised unit weight error'' (RUWE) to be $\le$ 10. These edits removed 5207 stars from the GCNS velocity sample, but did not appreciably change any of the gross kinematic statistics of this sample, such as in Table \ref{table:kinematic-stats}. 

At present, we do not include a model for the velocity of ISO ejection from their host system.
Most ISOs are thought to originate though ejection from stellar systems and thus will have their ejection velocities combined with  their host system galactic velocities. Any ejection process is likely to favor the production of objects with low ejection velocities, but bodies ejected from close to their host stars could conceivably have large outgoing v$_{\infty}$, which would spread out their velocity distributions \citep{Siraj-Loeb-2020-c} and could increase the high velocity tail in the ISO velocity distribution function. 

Once in galactic orbits ISOs are subsequently subject to the same gravitational perturbations as stars and, absent any significant drag or radiation pressure forces 
\citep{Eubanks-2019-b}, will share the dynamical modifications of those velocities by galactic tides and resonances \citep{Dehnen-2000-a}. Through sampling of ISO compositions, telescopically or by spacecraft exploration, it may be possible to distinguish between ISOs originating with a particular dynamical stream, and those originating elsewhere and gravitationally captured in that stream. 

\subsection{The GCNS Local Standard of Rest}

As a check on our treatment of the GCNS data, we estimated the kinematic Local Standard of Rest (LSR) for the GCNS velocity sample, the vector mean velocity of the sample relative to the Sun. (Note that this results in an  
estimate of  the LSR relative to the Sun, not the Sun relative to the LSR as is sometimes reported). We did this with  normal distribution fits to the galactic U, V and W velocity components for the velocity sample (where UVW are defined in a right-hand system with unit vectors pointing towards the galactic center, in the direction of galactic rotation, and towards the galactic North pole, respectively). The resulting LSR estimate 
is (-11.0$\pm$0.2, -15.4$\pm$0.2, -7.2$\pm$0.1) km s$^{-1}$  The GCNS LSR velocity is almost 4 km s$^{-1}$ larger than the 
average  used in \citep{Eubanks-2019-b}, with the difference being primarily in the V velocity component. 
 
A comparison with a set of independent estimates 
\citep{Schonrich-et-al-2009-a,Francis-Anderson-2014-a,Bland-Hawthorn-Gerhard-2016-a,Bobylev-Bajkova-2017-a} has a (U,V,W) component root mean square (rms) of (2.2, 4.1, 0.8) km s$^{-1}$, which can be taken as a reasonable estimate of the true uncertainty in these LSR components. The V velocity component is the least gaussian of the three galactic velocity components; this resulting larger uncertainty in the determination of the V of the LSR is also present in other LSR determinations
\citep{Gaia-et-al-2020-b,Francis-Anderson-2014-a,Schonrich-et-al-2009-a}. The GCNS LSR estimate is even closer to the 1I $\vec{v}_{\infty}$ vector, with a (U,V,W) 1I-LSR difference of (-0.54$\pm$0.18, -7.04$\pm$0.22, -0.62$\pm$0.13)  km s$^{-1}$ yielding a magnitude  $\vert\vec{\Delta\mathrm{v}}\vert$ = 7.1 km s$^{-1}$. The errors presented for this $\vec{\Delta\mathrm{v}}$ are formal errors; if the more realistic uncertainties provided above are applied the new 1I-LSR relative velocity difference is only of marginal statistical significance, supporting the young kinematic age estimated for that ISO \citep{Almeida-Fernandes-Rocha-Pinto-2018-a,Portegies-Zwart-et-al-2017-a}. 

\subsection{Stellar Velocity Probability Distribution Functions}

\begin{figure}[ht!]
\plotone{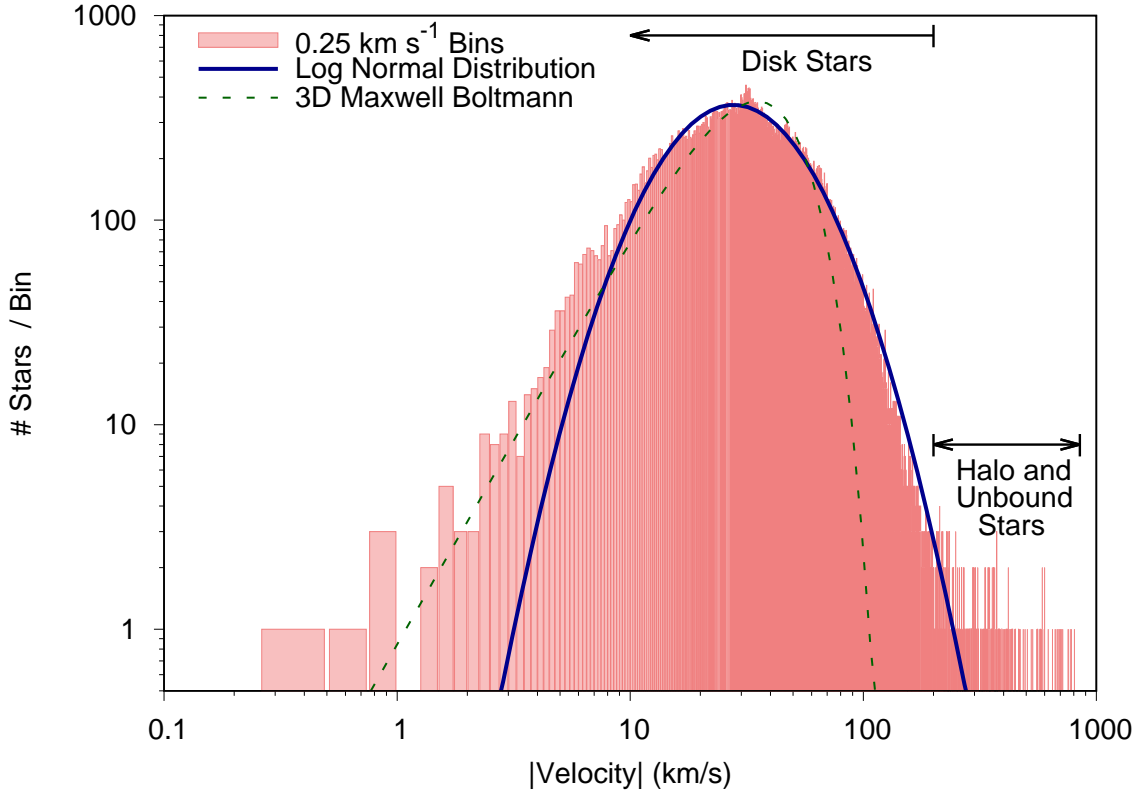}
\caption{Histogram of the distribution of 3-D GCNS stellar velocity magnitudes, $\vert$V$\vert$, for velocities relative to the solar system; basic kinematic statistics for these data are presented in Table \ref{table:kinematic-stats}. The solid and dashed curves show the 
Log-Normal distribution (Equation \ref{eq:LN-distro}) and the 
3-D Maxwell-Boltzmann distribution (Equation \ref{eq:MB-distro}), respectively, as fit to these data.  Halo stars are defined as having velocities $\ge$ 200 km s$^{-1}$ relative to the solar system, while candidate unbound stars have velocities
relative to the galactic barycenter, v$_{gal}$,
$\gtrsim$ 530 km s$^{-1}$. Halo and unbound stars are thus interspersed in this Figure. \label{Fig:Stellar-Distribution-1}
}
\end{figure}

Figure \ref{Fig:Stellar-Distribution-1} shows the histogram of the magnitude of the 3-D GCNS velocities, which peaks 
at $\sim$31.4 km s$^{-1}$. (A very similar distribution was also found, for transverse velocities only, by \citep{Amarante-et-al-2020-a}.)  
The GCNS velocity sample as a long high velocity tail, with 50\% of the stars having a velocity $\geq$ 40  km s$^{-1}$ and 5.2\% velocities $\geq$ 100  km s$^{-1}$  

\begin{figure}[ht!]
\plotone{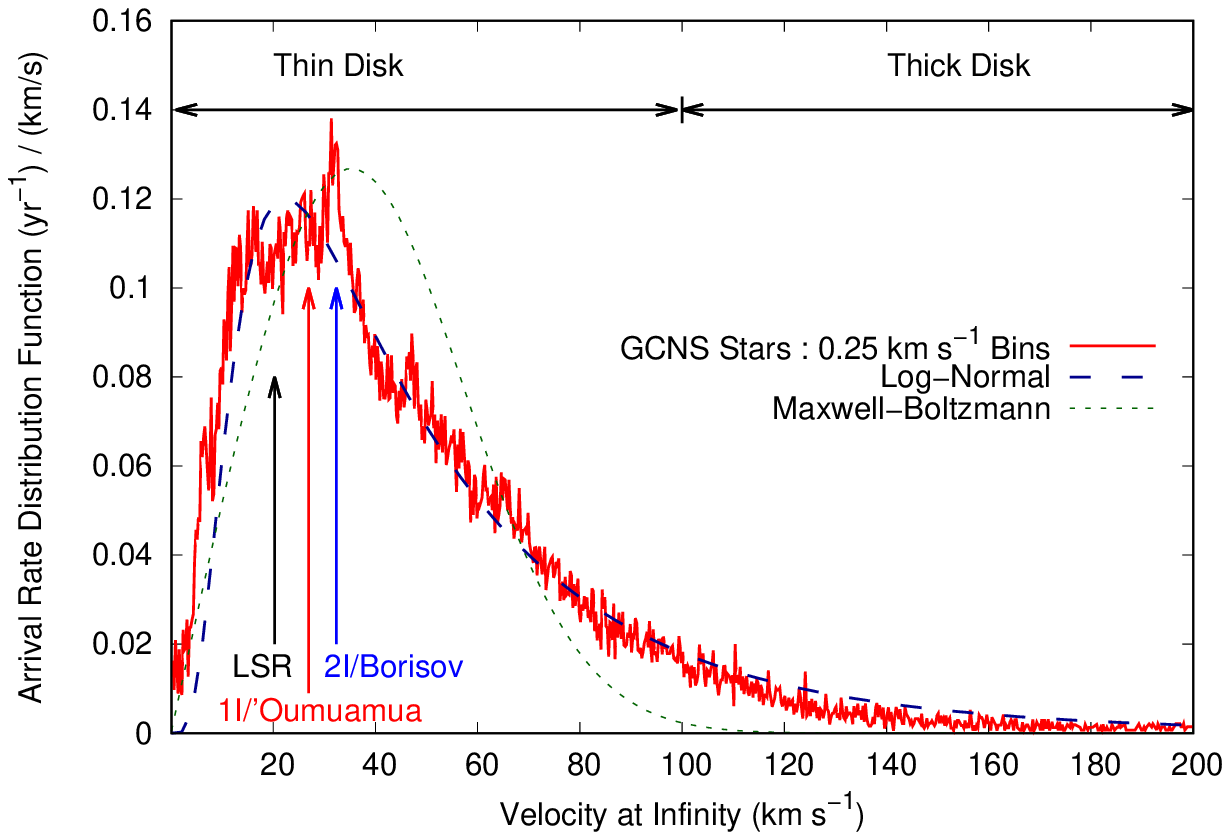}
\caption{The differential ISO arrival rate at the Earth's orbit, $\Gamma$(v$_{\infty}$,q = 1 AU), as derived from the the histogram-based GCNS distribution function shown in Figure \ref{Fig:Stellar-Distribution-1}, after accounting for gravitational focusing (Equation \ref{eq:diff-arrival-rate}). (The log-normal and 3-D Maxwell Boltzmann distributions have also had this model applied to them here.) Statistics of the cumulative integral of the GCNS kinematic model are provided in Table \ref{table:Gamma-stats}. The 
peak in the distribution at $\sim$31.4 km s$^{-1}$ is still present but has been noticeably  widened by gravitational focusing towards lower velocities. Even with gravitational focusing the median in the arrival flux distribution is at 38.0 km s$^{-1}$; half of the arriving ISOs will have velocities in the high velocity tail of the distribution. 
\label{fig:PDF-velocity}
}
\end{figure}

We investigated a number of distribution functions to model the 3D stellar velocity distributions.
The combined $\vert$V$\vert$ distribution of the GCNS is 
better described by a log-normal distribution than a 3-D Maxwell-Boltzmann distribution; we provide details of both such models here, and they are  shown in Figures \ref{Fig:Stellar-Distribution-1} and \ref{fig:PDF-velocity}. Our primary interest here is in estimating ISO arrival rates and we found it more useful to use the actual GCNS histograms, with typical
statistical uncertainties of order 1\%, for most of our analysis. 

\begin{figure}[ht!]
\plotone{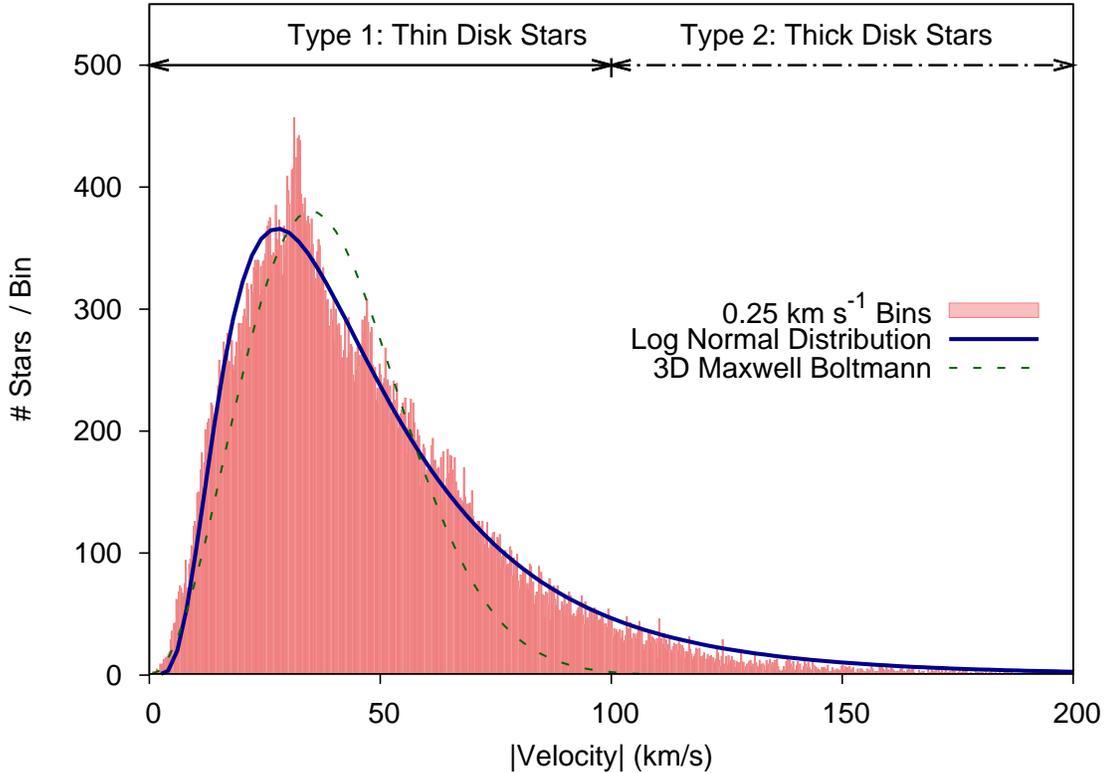}
\caption{The GCNS velocity sample histogram in Figure \ref{Fig:Stellar-Distribution-1} 
expanded to better show the velocity distribution for the thin and thick disk stars. 
of the distribution of 3-D GCNS stellar velocity magnitudes, $\vert$V$\vert$, for velocities relative to the solar system. As before, the solid and dashed curves show the 
Log-Normal and 
3-D Maxwell-Boltzmann distributions. 
 \label{Fig:Stellar-Distribution-2}
}
\end{figure}

In general, the Maxwell-Boltzmann distribution is a somewhat better model for the low velocity tail in the stellar kinematics, while the Log-Normal distribution is considerably better at describing the distribution of stars with velocities $\gtrsim$ 100 km s$^{-1}$. Neither distribution is adequate to include the relatively small number of Halo and unbound stars. or the very sharp peak in the distribution at $\sim$31.4 km s$^{-1}$ seen in Figure \ref{Fig:Stellar-Distribution-2}.

The log-normal (LN) distribution is described by 
\begin{equation}
\mathrm{p}_{\mathrm{LN}}(\mathrm{v})\ =\ \frac{1}{\mathrm{v}\ \sigma_{\mathrm{LN}}\ \sqrt{2\pi}}\ 
\exp\left({-\frac{(\ln\ \mathrm{v}\ -\ \mu)^{2}}{2\sigma_{\mathrm{LN}}^{2}}}\right)
\label{eq:LN-distro}
\end{equation}
where $\mu$ and $\sigma_{\mathrm{LN}}$ are solve-for parameters with $\sigma_{\mathrm{NL}}$ = 0.624 $\pm$ 0.002 and $\mu$ =  3.715 $\pm$ 0.003, and the three dimensional (3-D) Maxwell-Boltzmann distribution described by
\begin{equation}
\mathrm{p}_{\mathrm{MB}}(\mathrm{v})\ =\ \sqrt{\frac{2}{\pi}}\ 
\frac{\mathrm{v}^2\ \mathrm{e}^{-\mathrm{v}^2/(2\sigma_{\mathrm{MB}}^{2})}}{\sigma_{\mathrm{MB}}^{3}}
\label{eq:MB-distro}
\end{equation}
where v is the magnitude of the  velocity vector; the curve shown in Figure \ref{Fig:Stellar-Distribution-1} uses $\sigma_{\mathrm{MB}}$ = (26.14 $\pm$ 0.13) km s$^{-1}$.

\begin{table}[ht]
\centering
\caption{Kinematic statistics for the 77,132 stars in the ``3-D'' subset of the  GCNS catalog, the stars which have full 3-D velocity estimates and pass the astrometric and RUWE quality checks. 
Velocities are all relative to the Sun except for v$_{gal}$, which is the magnitude of the velocity relative to the galactic barycenter.  The kinematic division 
into thin disk, thick disk and Halo stars is that used in \citep{Gaia-et-al-2020-b} and used to define ISO velocity types by \cite{Hein-et-al-2020-b}. The median velocity of this integrated distribution is $\sim$40 km s$^{-1}$.
}
\begin{tabular}{c c c c}
\cline{1-4}
 Type  & Population &                  Velocity Range & Fraction  \\
    1  & Thin Disk  & 0 - 100  km s$^{-1}$            & 94.82\%              \\
Median &            &     40   km s$^{-1}$            & 50.00\%              \\
    2  & Thick Disk & 100 - 200 km s$^{-1}$           &  4.71\%              \\
    3  & Halo       & 200 - 679  km s$^{-1}$          &  0.44\%             \\
    4  & Unbound    & v$_{gal}$ $>$ 530  km s$^{-1}$  &  0.03\%             \\
    5  & Oort Spike & 0 - 1.5  km s$^{-1}$            &  0.009\%            \\
 \cline{1-4}
\end{tabular}
\label{table:kinematic-stats}
\end{table}

\subsection{The Local Consistency of the Stellar Number Distribution Function} 
\label{Subsec:Local-Consistency}

The stellar statistics and distribution functions derived from the GCNS are of course an average over a region of space with a diameter of 100 pc  and include stars moving towards and away from  the solar system. However, as the complete 6-dimension (6-D) position and velocity coordinates are available for the velocity subset of the GCNS used here, comparisons of regional subsets of the GCNS data are straightforward. We did this for 3 subsamples of these data, the close stars (with distances $\leq$ 50 pc), and the incoming and outgoing samples (those with radial velocities directed towards or away from the Sun, respectively).  (Note that the incoming and outgoing samples are mutually disjoint but together contain all of the GCNS velocity sample, while close sample is a subset of that sample.)

\begin{figure}[ht!]
\plotone{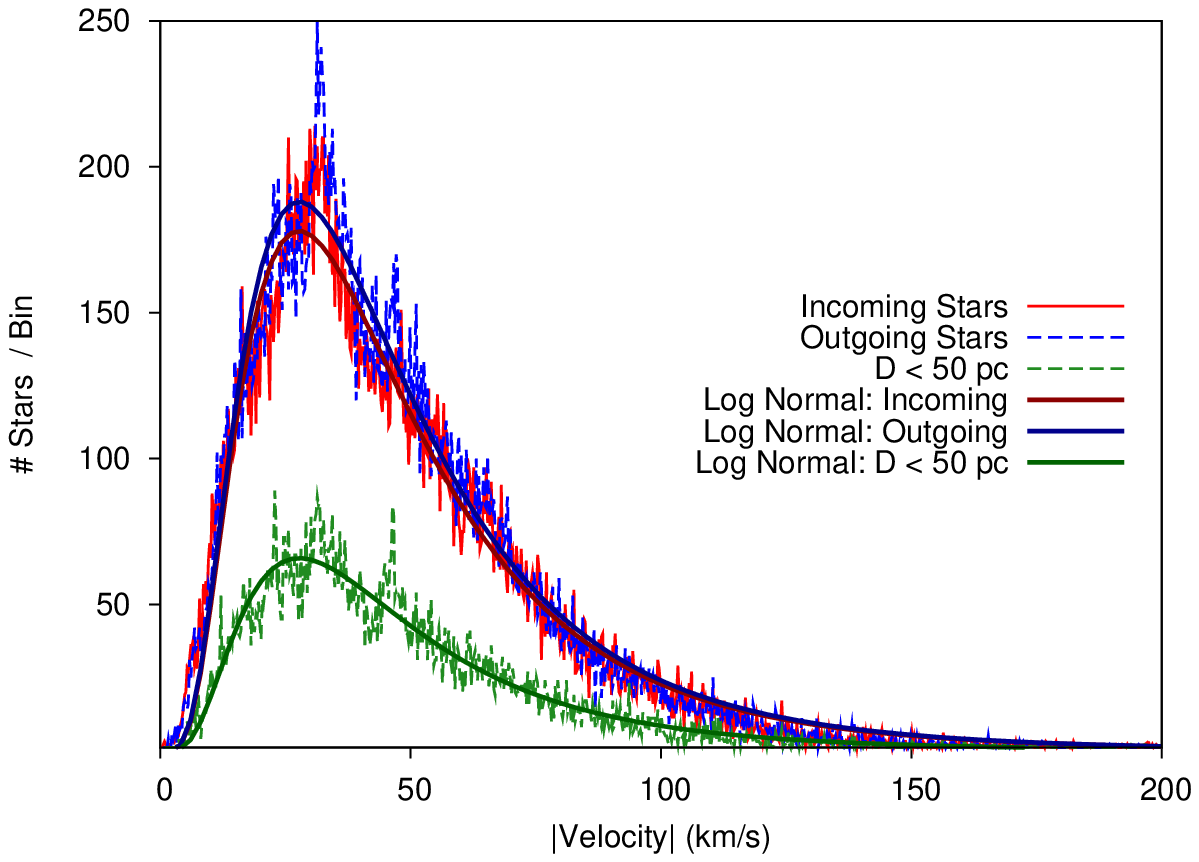}
\caption{The GCNS velocity sample histogram in Figure \ref{Fig:Stellar-Distribution-2} for the close, incoming and outgoing subsets of the 
velocity data as defined in subsection \ref{Subsec:Local-Consistency}. The solid curves show the 
Log-Normal  distribution described in 
Equation \ref{eq:LN-distro} scaled for each data subset, which in each case provides a reasonable fit to the subset histogram. 
The narrow peak in the distribution at $\sim$31.4 km s$^{-1}$ 
is only present in the outgoing stellar distribution function, and is thus not likely to be present in the ISO flux at the solar system. 
\label{fig:Local-PDFs}
}
\end{figure}

There is no statistically significant evidence for a change in either the median or the width of the log-normal distribution between the complete sample and any of these subsets. Figure \ref{fig:Local-PDFs} shows that the velocity distribution function of all three subsets are reasonably described by the Log-Normal  distribution given in Equation \ref{eq:LN-distro} scaled to match the number of stars in the subset. We conclude that the complete GCNS velocity sample reasonably reflects of kinematics of obkects in the solar neighborhood.

\section{ISO Arrival Rate Estimates} 
\label{Sec:Arrival-Rate-Estimates}

The differential arrival rate at the solar system for a given v$_{\infty}$ and q is the product of the volume sample rate $\gamma$, the number density n$_{\mathrm{ISO}}$ and the velocity distribution p$_{ISO}$(v), yielding (in our model)
\begin{equation}
\Gamma(\mathrm{v}_{\infty},\mathrm{q})\ =\ \mathrm{n}_{\mathrm{ISO}}\ 
\gamma(\mathrm{v}_{\infty},\mathrm{q})\ p_{\mathrm{GCNS}}(\mathrm{v}_{\infty}) .
\label{eq:diff-arrival-rate}
\end{equation}
$\Gamma(\mathrm{v}_{\infty},\mathrm{q})$, the expected differential arrival rate for incoming 1I-type ISOs, can be integrated to estimate the total arrival date for a given velocity range, and can be scaled linearly to account for different  $\mathrm{n}_{\mathrm{ISO}}$ for other objects, for example for more massive objects. 
Figure \ref{Sec:Stellar-Velocity-Distributions} shows 
estimates of the differential ISO arrival rate, 
$\Gamma(\mathrm{v}_{\infty},\mathrm{q}\ =\ 1\ \mathrm{AU})$, using Equation \ref{eq:diff-arrival-rate}, the data shown in Figure \ref{Fig:Stellar-Distribution-1} and the models in Equations \ref{eq:LN-distro} and \ref{eq:MB-distro}. 
Table \ref{table:Gamma-stats} provides summary statistics for $\Gamma$ at 1 AU. 
There is a broad peak in $\Gamma$ at $\sim$30 - $\sim$35 km s$^{-1}$; the v$_{\infty}$ for 2I, at 32.35 km s$^{-1}$, lies near the center of this peak. Half of the predicted ISO arrivals will have velocities $\geq$ 38 km s$^{-1}$, and 50\% of the arrivals are predicted to fall between 22.5 and 62.5 km s$^{-1}$.

\begin{table}[ht]
\centering
\caption{Integrated Flyby Rates, $\int\Gamma$,  estimated for 1I type ISOs from Equation \ref{eq:diff-arrival-rate} and the GCNS data in 
Figure \ref{Fig:Stellar-Distribution-1}.
As in Table \ref{table:kinematic-stats}, velocities are all relative to the Sun except for type 4 unbound objects. 
 The estimate for unbound ISOs is described 
in Subsection \ref{Extra-Galactic}. Fractions are compared to the total  
estimate of 6.90 ISOs / yr integrated over all velocities. Note that the midpoint of the velocity distribution, at  38.0 km s$^{-1}$, is hardly changed by gravitational focusing from the value in Table \ref{table:kinematic-stats}.
}
\begin{tabular}{c c c r}
\cline{1-4}
 Type &  Velocity Range &    $\int\Gamma$  & \multicolumn{1}{c}{Fraction}\\
    1 &   0 - 100 km s$^{-1}$              & 6.34 / yr  & 91.9\%  \\
Median  & 38.0 km s$^{-1}$              & 3.45 / yr  & 50\%    \\
    2 &   100 - 200 km s$^{-1}$            & 0.44 / yr  & 6.4\%  \\
    3 &   $>$ 200 km s$^{-1}$              & 0.09 / yr  & 1.3\%  \\ 
    4 &   v$_{gal}$ $\ge$ 530 km s$^{-1}$  & 0.03 / yr  & 0.4\%  \\
    5 &   0 - 1.5 km s$^{-1}$              & 0.01 / yr  & 0.2\%  \\
 \cline{1-4}
\end{tabular}
\label{table:Gamma-stats}
\end{table}

\subsection{Arrival Rates of 2I/Borisov Type Interstellar Objects}

2I/Borisov was discovered on
August 30, 2019, (M.P.E.C. 2019-R106)  at an R magnitude of 17.8 and a distance of $\sim$3.72 AU from the Earth and $\sim$2.99 AU from the Sun; precovery data was later found 
back to December 2018, when it was 7.8 AU from the Sun \citep{Ye-et-al-2019-b}.
It passed through perihelion at $\sim$2.01 AU from the Sun on December 8, 2019, 
reaching a peak apparent magnitude of about 15. Clearly, 2I was a much brighter and easier to detect object than 1I. Based on data from the IAU Minor Planet Center database, long period comets have been routinely discovered at similar magnitudes since at least Comet Kohoutek (C/1973 E1) in 1973 \citep{Eubanks-2019-b}, and have been reported as far back as 1955 \citep{Sekanina-2019-b}. Since even a 2I sized comet would be quite noticeable if it passed near the Earth, and since the orbit of a hyperbolic comet could potentially have been recognized, given sufficient data, as far back as the time of Edmund Halley, it seems clear that the space density of 2I type InterStellar Comets,  n$_{\mathrm{ISC}}$,  is considerably less than the 1I n$_{\mathrm{ISO}}$.

The number density estimate for 2I/Borisov type interstellar comets can be estimated from the discovery rate for similar long period comets. Since about the year 2000, 
there has been a considerable increase in the rate of discovery of long period comets, from 4.2 yr$^{-1}$  in the 20th century to 27.1 yr$^{-1}$  in the first 17 years of the 21st century 
\citep{Krolikowska-Dybczynski-2019-a}. This increase is largely in the discovery of objects with q $\geq$ 3.1 AU, which rose from $\sim$0.8 yr$^{-1}$ in the 20th century to 11.3 yr$^{-1}$ for the first part of the 21st century. 

We therefore modeled 
n$_{\mathrm{ISC}}$ assuming that a 2I type object would have been detected if one had arrived with q $\leq$ 2 AU for the last 50 years, and q $\leq$ 3 AU for the last 20 years. Using these maximum perhelion distances, and assuming that  2I comets follow the stellar velocity distribution in Figure \ref{Fig:Stellar-Distribution-1},
results in a number density estimate for 2I-type interstellar comets of
\begin{equation}
\mathrm{n}_{\mathrm{ISC}}\ \lesssim\ 7.2\ \times\ 10^{-5}\ \mathrm{AU}^{-3}\ .
\label{eq:ISC-number-density}
\end{equation}
This estimate is $\sim$ a factor of two smaller the estimate of 1.4 $\times$ 10$^{-4}$ AU$^{-3}$ 
derived by \cite{Engelhardt-et-al-2017-a} before the discovery of either 1I or 2I. 

\begin{figure}[ht!]
\plotone{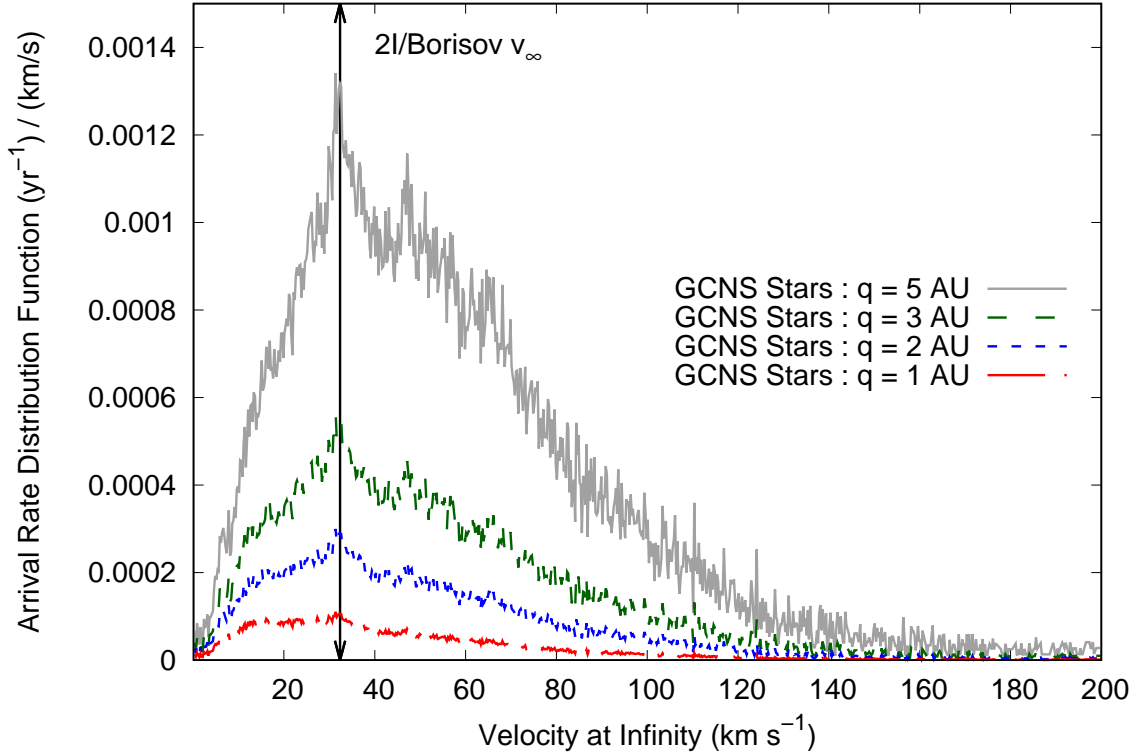}
\caption{The ISO differential arrival rate, $\Gamma$, for different values of the perihelion of the catchment orbit, as in Figure \ref{fig:PDF-velocity} but using the n$_{ISO}$ derived  for 2I/Borisov type objects. The observed 2I v$_{\infty}$ is very close to the peak in the $\Gamma$ distribution predicted from the EDR3 GCNS data. Note as the perihelion distance increases, the effect of gravitational focusing on the estimate velocity distribution decreases, which as a result pushes the arrival distribution towards higher velocities. The median velocity for arrivals with a 5 AU perihelion, for example, is 51 km s$^{-1}$, versus 38 km s$^{-1}$ for arrivals at 1 AU. 
\label{fig:2I-velocity}
}
\end{figure}

Figure \ref{fig:2I-velocity} shows the differential arrival $\Gamma_{\mathrm{ISC}}$ estimates using the GCNS data. Kinematically, 2I appears to be a typical interstellar comets, with the 2I v$_{\infty}$ being almost exactly at the maximum of $\Gamma_{\mathrm{ISC}}$.
Table \ref{table:Comet-Gamma-stats} provides arrival predictions based on Equations \ref{eq:cross-section} and \ref{eq:ISC-number-density}; the completeness to q = 5 AU is intended to represent the survey efficiency of the Vera Rubin Observatory. 
Even with forthcoming increases in survey sensitivity we predict the discovery of less than 1 2I-type ISO per decade, and most of these can be expected to have large perihelia, increasing the difficulty of interceptor missions 
\citep{Schwamb-et-al-2020-a} for interstellar comets. 

The ratio of the derived 2I and 1I number densities, n$_{\mathrm{ISC}}$ / n$_{\mathrm{ISC}}$. is roughly 7 $\times$ 10$^{-4}$. Assuming effective diameters of 1.4 and 0.065 km for these two objects, the size distribution dn/dD is $\propto$ D$^{-2.4}$. Given that this based on only two objects with considerable uncertainties in both n and D, this estimate should be regarded as suggestive only.

\begin{table}[ht]
\centering
\caption{Integrated Flyby Rates, $\int\Gamma$, for 2I-size Interstellar Comets,  estimated  from Equation \ref{eq:diff-arrival-rate} and Eq. \ref{eq:ISC-number-density}
and the GCNS data in 
Figure \ref{Fig:Stellar-Distribution-1}. Note that these rates are per century.
}
\begin{tabular}{c c c}
\cline{1-3}
 \multicolumn{1}{c}{Orbit} &  Description     &  $\int\Gamma$  \\
  q$_{p}$ = 1 AU & All v$_{\infty}$ & \multicolumn{1}{c}{0.5 / cy}\\ 
  q$_{p}$ = 2 AU & All v$_{\infty}$ & \multicolumn{1}{c}{1.4 / cy}\\ 
  q$_{p}$ = 3 AU & All v$_{\infty}$ & \multicolumn{1}{c}{2.8 / cy}\\ 
  q$_{p}$ = 5 AU & All v$_{\infty}$ & \multicolumn{1}{c}{7.0 / cy}\\ 
  q$_{p}$ = 5 AU & v$_{\infty}$ $\le$ 1.5 km s$^{-1}$ & {0.005 / cy} \\
 \cline{1-3}
\end{tabular}
\label{table:Comet-Gamma-stats}
\end{table}

\section{Science With Extreme Hypervelocity Impacts} 
\label{App:Impact-Science}

\cite{Hein-et-al-2020-b} discuss various missions types for ISO exploration, including fast flyby missions, 
sample return missions (a fast flyby through the coma of a interstellar comet, or an artificial coma formed by impacts on an interstellar asteroid) and 
rendezvous missions (where the spacecraft and ISO match velocities, possibly including orbiting or even landing on the interstellar body). 
No matter what technique is used to reach an passing ISO, the relative spacecraft-ISO velocity at the time of their encounter may be large. 
ISO velocities at a given distance, R, from the Sun, v(R,v$_{\infty}$), are given by
\begin{equation}
\mathrm{v}(\mathrm{R},\mathrm{v}_{\infty})\ = \sqrt{\mathrm{v}_{\infty}^2 + \mathrm{v}_{esc}(\mathrm{R})^2} ,
\label{eq:ISO-velocity}
\end{equation}
ignoring planetary perturbations and any other forces. (Note that v(R) depends only on the current distance from the Sun, and not the perihelion distance.)  
An interstellar object with a negligible v$_{\infty}$ will have a heliocentric velocity of 41.2 km s$^{-1}$ at 1 AU, and a velocity relative to the Earth ranging between 12.3 and 71.9 km s$^{-1}$, depending on the relative orientation of the orbital velocities. At the median ISO v$_{\infty}$ of 38 km s$^{-1}$, the corresponding velocity range at 1 AU is 26.9 to 86.5 km s$^{-1}$.

Hypervelocity impacts are defined to have relative velocities  $\gtrsim$ 3 - 5  km s$^{-1}$,  speeds where the strength of materials is negligible compared to impact forces. Biomarkers can survive at least at the lower part of the hypervelocity range \citep{Burchell-et-al-2014-a}, and it seems likely that a immediate (or prompt) cloud of very hot plasma would eject cooler, chemically intact, material that could be sampled in a fast flyby. Some material can survive even very fast impacts \citep{McDermott-et-al-2016-a}, and it seems possible that even hypervelocity ISOs could serve as a means for lithopanspermia \citep{Belbruno-et-al-2012-a}. 
At even higher velocities,  the impact energy / atom controls the prompt response to the impact. As an example, while it  only takes about 0.4 eV / molecule to vaporize water, water molecules have a binding energy of $\sim$4.4 eV, and the first ionization of both Hydrogen and Oxygen requires $\sim$13.6 eV.  There is thus a profound difference between a hypervelocity impact at 5 km s$^{-1}$ (roughly 2 eV / water molecule), which will produce mostly superheated steam, and an impact at 15 km s$^{-1}$ (roughly 18 eV / water molecule), which will produce an ionized plasma. Impacts with energies / atom $\gtrsim$ 20 eV can thus be usefully described as extreme hypervelocity impacts, and ISO impact experiments will alnost entirely be extreme hypervelocity impacts. The resulting ionized prompt plumes will  producing radiation at Extreme UltraViolet (EUV, 10 to 120 nm) and soft X-ray (0.1 to 10 nm) and even hard X-ray wavelengths ($\leq$0.1 nm), depending on the collision energy, which can be used to investigate the physics of the impact and the composition of the impacted bodies.
Impacts at these velocities will strip off of multiple electron shells, creating highly ionized atoms and yielding prompt radiation radiation containing multiple electron recombination lines \citep{Eubanks-et-al-2020-b}, but will not be energetic enough to cause nuclear reactions.

\subsection{The Physics of Hypervelocity Impacts}
 
The hypervelocity impact technique was pioneered in 2005 by the Deep Impact (DI) mission, 
which struck the comet Tempel 1 with an impactor at an 
impact velocity of $\sim$10.3 km s$^{-1}$ \citep{AHearn-et-al-2005-a}. 
The DI impactor largely consisted of a 178.4 kg copper mass. 
Here, we model impacts with a probe, assumed to be made of pure $^{65}$Cu to avoid contamination of the prompt plume spectra, and determine the energies reached for various atomic species as function of the impact velocity. 

\begin{figure}[ht!]
\plotone{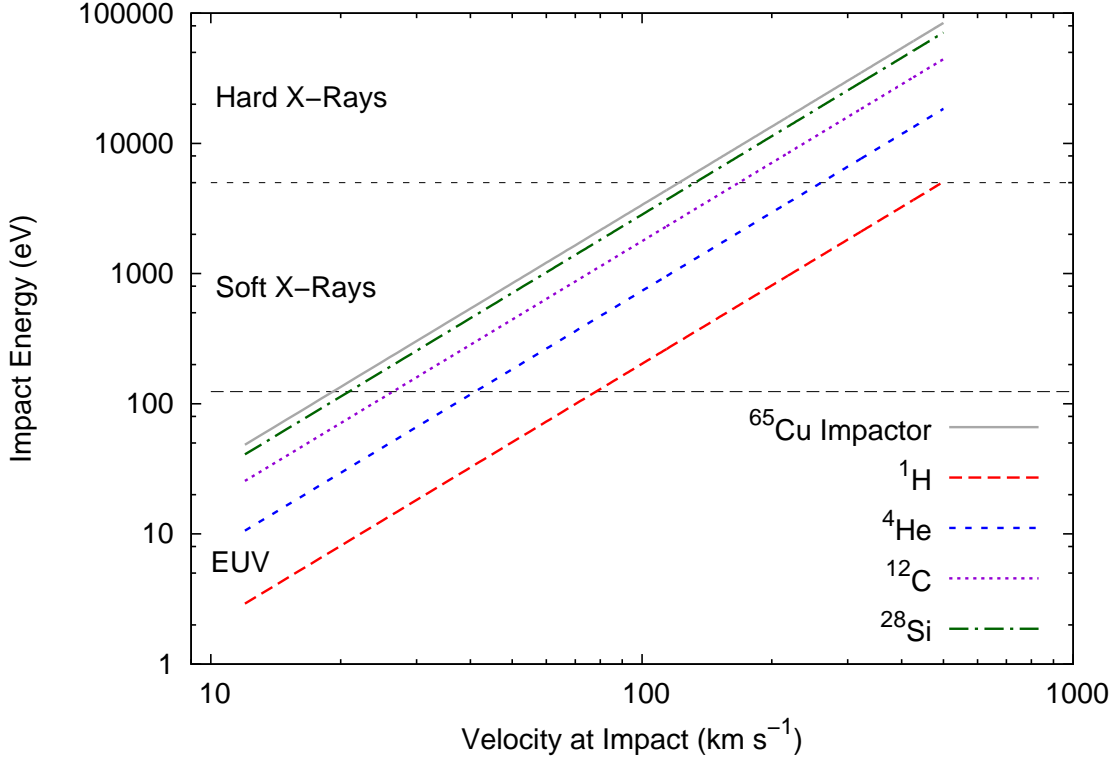}
\caption{Prompt impact energies as a function of the impact velocity for various atomic species, 
using Equation \ref{eq:elastic-1-D-A} assuming an impactor constructed entirely of isotopically pure Copper ($^{65} $Cu).
While the slower thin disk ISOs may predominantly produce extreme UltraViolot (EUV), 
thick disk, halo and unbound ISOs collisions will almost certainly be at $>$ 100 km s$^{-1}$ and thus should be be dominated by soft or even hard X-rays. 
\label{fig:Impact-Energies}
}
\end{figure}

A small impactor will not change the velocity of an impacted ISO by more than a few mm s$^{-1}$, and so a reference frame fixed in the the ISO can be viewed as an inertial frame, and the atomic constituents of the ISO can be viewed as initially at rest in that frame. Assume, as a first order approximation, a non-relativistic head-on elastic atomic collision between an atom in the impactor, of mass m$_{i}$ and initial velocity 
v$_{i_{in}}$, and an atom in the ISO, with mass m$_{\mathrm{ISO}}$ and zero velocity in the ISO rest frame. Then the post-collision velocities
in the ISO rest-frame  are given by
\begin{equation}
v_{i_{out}} = \frac{m_{i} - m_{\mathrm{ISO}}}{m_{i} + m_{\mathrm{ISO}}}\ v_{i_{in}}
\label{eq:elastic-1-D-i}
\end{equation}
and
\begin{equation}
v_{\mathrm{ISO}_{out}} = \frac{2\ m_{i}}{m_{i} + m_{\mathrm{ISO}}}\ v_{i_{in}}  .
\label{eq:elastic-1-D-A}
\end{equation}
Atoms with small atomic mass compared to the constituents of the impactor will receive a large velocity change (up to twice the impact velocity) but a relatively small fraction of the  incoming atom's Kinetic Energy (KE), while more massive atoms will have a smaller velocity change, but can absorb more of the incoming atom's KE. The energies considered in this paper are not large enough to initiate most nuclear reactions, but it is reasonable to assume that cohesive and molecular bonds will be broken, and electrons removed, up to the maximum amount of energy available. 
Figure \ref{fig:Impact-Energies} shows  impact energies from Equation \ref{eq:elastic-1-D-A} for Hydrogen, Helium, Carbon and Oxygen, common constituents in solar system comets and asteroids, assuming an impact by a $^{65}$Cu probe at the indicated impact velocity.

Although there is one impact velocity for any given impact, 
the different atomic masses of the various ISO constituents mean that
these atoms will gain different amounts of energy per collision, and thus will be at 
different temperatures. Once the prompt impact plasma forms, 
the temperatures will be rapidly equalized through equipartition of energy, which will increase the kinetic energy of light elements and decrease the energy of the heavier elements in a given composition. This warming of the light elements should be sufficient to produce the Lyman alpha transition for Hydrogen 
at 121.6 nm (10.2 eV) for almost any ISO fast flyby \citep{Eubanks-et-al-2020-b}. 
The prompt energies shown in Figure \ref{fig:Impact-Energies} are large enough for collisions at velocities 
$\geq$ 100 km s$^{-1}$ to general K-alpha X-ray spectral lines for many of the elements likely to be common in ISOs. Instruments such as the ALICE Ultraviolet Imaging Spectrograph, with sensitivity down to 52 nm (23.8 eV) 
\citep{Stern-et-al-2008-a} could be adopted to observe the ultraviolet spectra from ISO impacts but it will probably be necessary to develop special purpose X-ray telescopes to properly observe the full impact spectrum. 

\section{Discussion} 
\label{Sec:Discussion}

Our analysis indicates that incoming ISOs will include a substantial population of fast moving objects, which are likely to challenge both the surveys searching for ISOs and spacecraft missions intended to observe them. 
Asteroid surveys may have a streak limit, a limit on the acceptance of fast moving small objects, in order to reduce confusion with terrestrial satellites. 1I, at the time of its discovery, was moving at $\sim$6$\degree$ / day, close to, but  below, the then-current \textit{Pan-STARRS1} streak limit \citep{Do-et-al-2018-a}. Faster ISOs may conceivably been already observed by existing surveys,  ignored due to filters such as this, and be recoverable through re-processing \citep{Ye-et-al-2019-b,Robert-et-al-2021-a}. 
High dynamic range synthetic tracking (ST) offers a means of avoiding the streak limit in asteroid surveys, and of substantially improving the detectability of small fast-moving bodies \citep{Zhai-et-al-2018-a}; our analysis suggests that these surveys should try to accommodate as high an incoming velocity as possible, ideally
up to hundreds of km s$^{-1}$.

\subsection{Type 1 Objects: ISOs from the Thin Disk}
\label{Thin-Disk}

Galactic velocity dispersions tend to increase with time,
which can be used to estimate ``kinematic ages'' and was used to conclude that
1I is a relatively young object, with a kinematic age of $\sim$0.20 to 0.45 Gyr  \citep{Almeida-Fernandes-Rocha-Pinto-2018-a,Hallatt-Wiegert-et-al-2019-a}. 
\cite{Siraj-Loeb-2020-c} discussed (using a simple gaussian kinematic model for stellar velocities)
the broadening of the ISO velocity distribution by the ISO velocity at ejection from their originating stellar system (the outgoing v$_{\infty}$), which could potentially be as large as 50 km $s^{-1}$ for eject of objects by a rapidly orbiting  M dwarf planet in its habitable zone. 
However, the numerical simulations of \cite{Napier-et-al-2021-a} (which apply to both capture and ejection of small bodies) indicate that most ejections by Jupiter and other the planets in the solar system  would be at outgoing v$_{\infty}$ velocities as low as 1 km s$^{-1}$, with the ejection efficiency declining as v$_{\infty}^{-6}$ for higher velocities. These outgoing velocity vectors would in most cases be random compared to the galactic velocity of the host system, and so will tend to simply increase the mean and dispersion of the resulting ISO velocity distribution in the galaxy. Unless there are mechanisms favoring large ejection velocities, it will take the discovery of a fairly large number of ISOs to statistically determine such a broadening of ISO velocity distribution function.   

1I had a v$_{\infty}$ of only 26.4 km s$^{-1}$, close to the LSR velocity relative to the Sun, and smaller than the median velocity expected for incoming ISO. While 1I's velocity is not statistically unusual (50\% of incoming ISOs should have 22.5 $\leq$ v$_{\infty}$ $\leq$ 62.5), and it may be simply indicative of a relatively young ISO, it suggests that small ISOs may have a small enough mass-to-area ratio to be subject to drag by the Interstellar Medium, as was hypothesized for 1I due to its anomalous acceleration in the solar system \citep{Bialy-Loeb-2018-a,Eubanks-2019-b}. If these low mass-to-area ratios are in fact common, the small ISO velocity distribution will show a peak, not at the stellar velocity distribution peak, but near the LSR velocity. Even a small number of additional 1I-type ISO discoveries should begin to show whether this hypothesis is correct. 

\subsection{Type 2 Objects: Thick Disk ISOs}
\label{Thick-Disk}

The stars in the thick disk are older than the thin disk stars; stars in the solar neighborhood with ages $>$ 8 billion years are almost exclusively thick disk stars, while the thin disk stars are younger, and include some stars with low metallicity \citep{Haywood-et-al-2013-a}. Thick disk stars also typically have higher inclinations and higher eccentricities and thus higher velocity dispersion in their galactic orbits than thin disk stars. 
\citep{Recio-Blanco-et-al-2014-a}. Although it is reasonable to expect that there will be fewer thick disk ISOs arriving in the solar system per unit time, observing them will provide information about protoplanet formation in the early stages of galactic history.

\subsection{Type 3 Objects: Halo ISOs}
\label{Halo}

The galactic halo is an extended, roughly spherical component of the Milky Way galaxy, thought to contain about 1\% of the stars in the Milky Way distributed over a much larger volume. The halo includes stars (and, thus, presumably ISOs) collected as debris from past accretion events; material from the halo thus could provide signatures from the smaller galaxies destroyed in the past and constraints on the accretion history of the galaxy \citep{Johnston-2016-a}. The halo also contains stars (possibly very old) that formed \textit{in situ}, and ones that were ``kicked out'' of the galactic disk; all of this material could in principle be sampled through the discovery of halo ISOs.

\subsection{Type 4 Objects: Unbound ISOs}
\label{Extra-Galactic}

The star with the highest relative velocity in the GCNS 3D velocity dataset,  EDR3 
6814962601568904576 or L 714-88, has a velocity relative to the Sun of 805 km s$^{-1}$, and a velocity relative to the galactic center of 720 km s$^{-1}$, indicating that it is not bound to the Milky Way galaxy. A total of 24 stars in the GCNS 3D velocity dataset are unbound object candidates with galactocentric velocity estimates $\ge$ the galactic escape velocity, $\sim$530 km s$^{-1}$ in the solar neighborhood \citep{Marchetti-2020-a}.
As the galactic rotation is $\sim$238 km s$^{-1}$ at the Sun's galactic radius, no unbound star can have a velocity relative to the solar system smaller than $\sim$292 km s$^{-1}$. The apparently  unbound subset of the GCNS data is faster moving, with solar system velocities ranging between 419 and 777 km s$^{-1}$, while the maximum velocity relative to the solar system of a  bound GCNS object is 600 km  s$^{-1}$. \cite{Marchetti-2020-a} used the entire Gaia EDR3 velocity dataset ($\sim$7 million stars) and found 99 candidates with a probability $>$ 50\% to be unbound stars, or  
$\sim$ 10$^{-5}$ of their data set, roughly one order of magnitude below the GCNS estimate here. It is even possible that there is a population of  ``hypervelocity'' stars, and thus possibly ISOs, with galactic velocities $\geq$ 1000  km s$^{-1}$ \citep{Lingam-Loeb-2020-c}.

This suggests that there is at the solar system a flux of unbound ISOs, with a space density of order (10$^{-6}$ to 3 $\times$ 10$^{-5}$)  AU$^{-3}$, consisting either of objects ejected from our galaxy (probably from star-formation regions) or are arriving from other galaxies. Equation \ref{eq:cross-section} shows that the cross section of the solar system is amplified for such high velocity objects, with $\sigma\ \sim$ 400 AU$^{3}$ yr$^{-1}$, yielding a very approximate estimate of order 1 such object per century passing within the Earth's orbit. While it will be hard to detect such fast moving objects, and even a fast flyby may be beyond current technology (an object moving at 530 km s$^{-1}$ will traverse 1 AU in 3.3 days), they offer the potential of sampling extra-galactic and hyper velocity material, rendering then of considerable scientific interest, and motivating improvements in searches to find them and means to better explore them if they can be found.

\subsection{Type 5 Objects: Low Velocity ISOs in the Solar System}

The model estimate here for n$_{\mathrm{ISC}}$ indicated that true interstellar comets with incoming  v$_{\infty}$ $\leq$ 1.5 km s$^{-1}$ should have an almost negligible arrival rate, much less than 1 per millennium even for improved future survey depths. As these same survey improvements should result in the discovery of dozens of Oort spike comets per year, it seems clear that any apparent interstellar comets with these low velocities will likely be from the Oort cloud of the solar system, either Oort cloud objects that are bound to the solar system but whose orbits appear to be unbounded due to unmodeled perturbations, or unbound objects previously lost but now re-encountering the Sun.

The Oort cloud of our solar system is thought to presently contain as many as 5 $\times$ 10$^{11}$ comets with diameters of $\sim$1 km or smaller spread up to 
10$^{4}$ to 10$^{5}$ AU from the Sun  \citep{Francis-2005-a}.
Many of these objects must be lost over time due to perturbations from galactic tides and passing objects. As the orbital velocity about the Sun at a semi-major axis of 1 light year is only about 100 m s$^{-1}$, unbound Oort cloud objects will drift very slowly away from the Sun, at order 10$^{-7}$ c, and can then drift back towards the solar system under the influence of galactic gravitational tides and passing stars. 

Analysis of stellar binaries indicates that in the galactic disk wide binary stars that become unbound will drift slowly apart, and can be recognized through common velocities at separations of up to 20 pc \citep{Kamdar-et-al-2019-a}.
There is no reason not to expect the orbits of unbound comets not to evolve similarly
\citep{Correa-Otto-Calandra-2019-a}, forming an extended volume of unbound Oort cloud objects (an ``Oort group'')
drifting away over durations up to 10$^{8}$ years, i.e., over intervals comparable to the galactic rotation period. If order 10$^{11}$ Oort cloud objects are assumed to be lost per Gigayear (so that roughly half of the Oort cloud has been lost in the history of the solar system), then at any one time the Sun would be surrounded by an accumulation of roughly 10$^{10}$ \textit{unbound} objects moving slowly away from the solar system. The bound Oort cloud will thus be surrounded by a physically larger unbound Oort group containing objects that have escaped the solar system's gravity but have not yet moved away. Objects in this unbound group can be perturbed by galactic tides or passing massive objects to move closer to the Sun, and if they enter the inner solar system it will almost certainly be with a very low v$_{\infty}$.

The recent study of \cite{Napier-et-al-2021-a} indicates that as v$_{\infty}$ goes to zero the Sun (or any star) is increasing likely to gravitationally capture the incoming object, through the slow motion of the Sun with respect to the solar system barycenter. It appears that  order 50\% of all  incoming objects with v$_{\infty}$ $\leq$ 50 m s$^{-1}$ will be captured by this mechanism. These extremely low velocity objects are easily captured but will be also easily lost from the solar system, either at aphelion, through perturbations, or at the next perihelion passage. This suggests that Type 5 ISOs are likely to be either Oort cloud or Oort group objects, and also that some of these object may  be lost and captured multiple times over the history of the solar system. This hypothesis can of course be tested directed by isotopic analysis of Type 5 ISOs once these are found.

\section{Conclusions}
\label{Sec:Conclusions}

Interstellar objects likely formed very far from the solar system
in both time and space; their direct exploration will constrain the formation and history of small bodies, situating them within the dynamical assembly and chemical evolution of the Galaxy.
The velocities of many of these objects in the solar system will make their detection and their \textit{in situ} exploration by spacecraft challenging, but not impossible with present and near future technology.

If the number density estimates based on the discovery of 1I/'Oumuamua are even approximately correct, there should be a number of 1I-type interstellar objects in the solar system at any one time, mostly with considerably higher 
velocities at infinity than 1I had. Current \citep{Do-et-al-2018-a,Ye-et-al-2019-b} and near future \citep{Seaman-et-al-2018-a} sky surveys should start finding a regular stream of 1I-type ISOs in the near future. If so, the prospects for a near term ISO mission in the inner solar system will become brighter, and it should be possible to rapidly determine the mass-to-area ratios for this class of objects. Even a fast flyby of an ISO passing through the solar system would be scientifically very rewarding, especially if the object can be subjected to the analysis of impact signatures. 

Unfortunately, 2I-type interstellar comets appear to be much rarer, which a much smaller infall rate, and will probably be  a decadal phenomena. Of interest, of course, is the existence and number density of intermediate, sub-km sized, ISOs, which could be either asteroidal or cometary in nature. These objects should exist, and should plausibly have an intermediate infall-rate between 1I and 2I-type objects. The discovery of even a few intermediate mass objects would substantially improve our knowledge of their number density mass-spectrum, which will be important for determining how ISOs form in the galaxy.

For decades to come, ISOs (including the already discovered 1I and 2I and any that are likely to be discovered) will be substantially easier to explore than any nearby stellar system. 
A long term program to find and explore ISOs will initiate the direct exploration of bodies beyond the solar system, can begin now with current technology, and will both assist and be assisted by the development of spacecraft and instrument technologies for interstellar travel. \cite{Hibberd-Hein-2020-a} demonstrate that a mission to 'Oumuamua would be feasible, using a GW-scale beaming infrastructure and a series of 1-100 kg probes. Any of the proposed technologies  being developed for interstellar flight or
for missions to the gravitational lens foci of the Sun \citep{Lubin-2016-a}, such as small sailcraft \citep{Turyshev-et-al-2020-a} and power beamed ``chipsats''   \citep{Hein-et-al-2017-b} could be used to explore ISOs. Even as propulsion technology is being developed to directly explore incoming and outgoing ISOs, new technology and new instruments will be required to best find passing ISOs, and to best utilize the opportunities that new ISO discoveries will make possible.

\begin{acknowledgments}
TME acknowledges support provided by Space Initiatives Inc and S.A. Eubanks. 
ML acknowledges support provided by the Florida Institute of Technology.
\end{acknowledgments}

\bibliographystyle{aasjournal}

\end{document}